\documentstyle[preprint,aps]{revtex}
\input epsf.tex
\def\DESepsf(#1 width #2){\epsfxsize=#2 \epsfbox{#1}}

\begin{document}
\preprint{\vbox{\hbox{CTP-TAMU-43-99}\hbox{COLO-HEP-439}\hbox{YUMS 00-10}\hbox{hep-ph/9911263}}} \draft
\title {Charmless hadronic B decays and the recent CLEO data}
\author{B. Dutta$^1$\footnote{b-dutta@rainbow.physics.tamu.edu} and Sechul Oh$^2$\footnote{scoh@kimcs.yonsei.ac.kr} }
\address{$^1$  Center For Theoretical Physics, Department of Physics, Texas
A$\&$M University, \\  College Station, TX 77843-4242, USA \\
 $^2$ Department of Physics and IPAP, Yonsei University, Seoul, 120-749, Korea}
\maketitle
\begin{abstract}
In the light of recent experimental data from the CLEO Collaboration we study the decays of $B$ mesons to a pair
of pseudoscalar ($P$) mesons, and  a vector ($V$) meson and a pseudoscalar meson, in the framework of
factorization. In order to obtain the best fit for the recent CLEO data, we critically examine the values of
several input parameters to which the predictions  are sensitive.  These input parameters are the form factors,
the strange quark mass, $\xi \equiv 1/ N_c$ ($N_c$ is the effective number of color), the CKM matrix elements and
in particular, the weak phase $\gamma$.   It is possible to give a satisfactory account of the recent experimental
results in $B \rightarrow PP$ and $VP$ decays, with constrained values of a \emph{single} $\xi$.  We identify the
decay modes in which CP asymmetries are expected to be large.
\end {abstract}

\newpage
\tightenlines
\section{INTRODUCTION}
The CLEO Collaboration\cite{1,2,3,4,5} has recently reported new
experimental results on branching ratios (BRs) of a number of
exclusive decay modes where $B$ decays into a pair of
pseudoscalars ($P$), a vector ($V$) and a pseudoscalar meson, or a
pair of vector mesons.   Several decay modes have been observed
for the first time,  such as $ B \rightarrow \pi^+ \pi^-$, $K^0
\pi^0$, $\omega \pi^{\pm}$, $\rho^0 \pi^{\pm}$, $\rho^{\pm}
\pi^{\mp}$, $K^{*\pm} \pi^{\mp}$, and $K^* \eta$.   Improved new
bounds have been put on the branching ratios for various modes,
such as $B \rightarrow K^{\pm} \pi^{\mp}$, $K^0 \pi^{\pm}$,
$K^{\pm} \pi^0$, $\omega K$, $\omega h^{\pm}$, and $K
\eta^{\prime}$.  A search for CP asymmetries in $B \rightarrow
K^{\pm} \pi^{\mp}$, $K^{\pm} \pi^0$, $K^0_S \pi^{\pm}$, $K
\eta^{\prime}$, and $\omega \pi^{\pm}$ has also been performed.

Recently, two works have been done to explain the recent CLEO
results for $B$ decays: one of them is based on the flavor SU(3)
symmetry \cite{8}, and the other one involves the
 framework of factorization \cite{9}.  In the latter work, the existence of
\emph{two different} effective numbers of color, $N^{{\rm eff}}_c (LL)$ and
$N^{{\rm eff}}_c (LR)$,  is essential, and their favored values are $N^{{\rm
eff}}_c (LL) =2$ and $N^{{\rm eff}}_c (LR) =6$ to
 explain the experimental data, except for $ \bar B^0 \rightarrow K^{*-} \pi^+$
and $\bar K^0 \pi^0$.

In the light of the recent CLEO data, in this work, we will analyze \emph{both} $B \rightarrow PP$ and $B
\rightarrow VP$ with a \emph{single} parameter $\xi$ in the framework of generalized factorization,  in order to
find satisfactory explanation compatible with all the recent experimental results.    Our approach will be
different from that of Ref. \cite{9}. Similar to the case of $B \rightarrow D$ (i.e., heavy $\rightarrow$ heavy)
decays, we will  assume only one \emph{universal} $\xi \equiv 1/ N_c$ ($N_c$ is the effective number of color) in
the analysis of both $B \rightarrow PP$ and $B \rightarrow VP$ ($P$ and $V$ are \emph{light} mesons). It has been
shown that in $B \rightarrow D$ decays a single $\xi$ can satisfactorily  explain experimental data for both $B
\rightarrow PP$ (such as $D \pi$) and $B \rightarrow VP$  (such as $D \rho$) \cite{10}. In order to achieve  this
goal, the values of all the input parameters, e.g., the  form factors, the strange quark mass, $N_c$, the
Cabibbo-Kobayashi-Maskawa (CKM) matrix elements etc.,  will be carefully examined  and constraints on these
parameters will be investigated.    We shall see that it is indeed possible to account satisfactorily for all the
recent experimental  data within our framework.   In addition, we will discuss CP asymmetries in the $B$ decays
and  identify the decay modes where the CP asymmetries are possibly large.  In our previous works \cite{6,7}, we
have shown that the predictions for $B \rightarrow PP$ and $VP$ modes  are sensitive to several input parameters,
such as the form factors, the QCD scale, the parameter $\xi \equiv 1/ N_c$, the CKM matrix elements, and the light
quark masses, in particular the strange quark mass.    With the improved recent data, the results (e.g., $\xi \sim
0$) obtained in the previous works are unlikely to be compatible with the experimental results.   Our main
emphasis was to explain the large branching ratio for $B \rightarrow K \eta^{\prime}$ within factorization.   In
order to explain the large branching ratio for $B \rightarrow K \eta^{\prime}$, different assumptions have been
proposed, e.g., large form factors \cite{11}, the QCD anomaly effect \cite{11a,11b}, high charm content in
$\eta^{\prime}$ \cite{12,12A,12a}, a new mechanism in the Standard Model \cite{12b}, or new physics like
supersymmetry without R-parity \cite{13}.   In this work we will assume that a certain (unknown) mechanism is (at
least in part) responsible for the large branching ratio for $B \rightarrow K \eta^{\prime}$ and hence will not
consider the decay modes having $\eta^{(\prime)}$ in the final states.

We organize this work as follows.  In Sec. II we discuss the
effective Hamiltonian and obtain the effective Wilson coefficients
at the scale $m_b$ for both $b \rightarrow s$ and $b \rightarrow
d$ transitions.  In Sec. III we describe the parametrization of
the matrix elements and define the decay constants and the form
factors.  In Secs. IV and V the two body decays $B \rightarrow PP$
and $B \rightarrow VP$ are analyzed in the framework of
factorization.  The CP asymmetries in the $B$ decay modes are also
discussed.  Finally, in Sec. VI our results are summarized.

\section{DETERMINATION OF THE EFFECTIVE WILSON COEFFICIENTS}
The effective
weak Hamiltonian for hadronic $B$ decays can be written as
\begin{eqnarray}
 H_{\Delta B =1} &=& {4 G_{F} \over \sqrt{2}} \left[V_{ub}V^{*}_{uq} (c_1 O^{u}_1
+c_2 O^{u}_2)         + V_{cb}V^{*}_{cq} (c_1 O^{c}_1 +c_2
O^{c}_2)  - V_{tb}V^{*}_{tq} \sum_{i=3}^{12} c_{i} O_{i} \right]
\nonumber \\  &+& {\rm H.c.} ,
\end{eqnarray}   where $O_{i}$'s are defined as
\begin{eqnarray}  O^{f}_{1} &=& \bar q \gamma_{\mu} L f \bar f \gamma^{\mu} L b,
\ \   O^{f}_{2} = \bar q_{\alpha} \gamma_{\mu} L f_{\beta} \bar f_{\beta}
\gamma^{\mu} L b_{\alpha},    \nonumber \\   O_{3(5)} &=& \bar q \gamma_{\mu} L
b \Sigma \bar q^{\prime} \gamma^{\mu} L(R) q^{\prime},  \ \   O_{4(6)} = \bar
q_{\alpha} \gamma_{\mu} L b_{\beta} \Sigma \bar q^{\prime}_{\beta}  \gamma^{\mu}
L(R) q^{\prime}_{\alpha},   \nonumber \\                  O_{7(9)} &=& {3 \over
2} \bar q \gamma_{\mu} L b \Sigma e_{q^{\prime}} \bar q^{\prime}  \gamma^{\mu}
R(L) q^{\prime} , \ \  O_{8(10)} ={3 \over 2} \bar q_{\alpha} \gamma_{\mu} L
b_{\beta} \Sigma e_{q^{\prime}} \bar q^{\prime}_{\beta} \gamma^{\mu} R(L)
q^{\prime}_{\alpha} ,   \nonumber \\   O_{11} &=&{g_{s}\over{32\pi^2}}m_{b}\bar
q \sigma_{\mu \nu}RT_{a}bG_{a}^{\mu \nu} \;,\;\; O_{12} = {e\over{32\pi^2}}
m_{b}\bar q \sigma_{\mu \nu}R b  F^{\mu \nu} \;,
\end{eqnarray}    where $L(R) = (1 \mp \gamma_5)/2$, $f$ can be $u$ or $c$
quark, $q$ can be $d$ or $s$ quark,  and $q^{\prime}$ is summed
over $u$, $d$, $s$, and $c$ quarks. $\alpha$ and $\beta$ are  the
color indices.  $T^a$ is the SU(3) generator with the
normalization ${\rm Tr}(T^q T^b) = \delta^{ab}/2$.  $G^{\mu
\nu}_a$ and $F^{\mu \nu}$ are the gluon and photon field strength.
$c_i$'s  are the Wilson coefficients (WC's). $O_1$ and $O_2$ are
the tree level and QCD corrected operators.  $O_{3-6}$ are the
gluon induced strong penguin operators.  $O_{7-10}$ are the
electroweak penguin operators due to $\gamma$ and $Z$ exchange,
and the box diagrams at loop level.   In this work we shall take
into account the chromomagnetic operator $O_{11}$, but neglect the
extremely small contribution from $O_{12}$. The dipole
contribution is in general quite small, and is of the order of
$10\%$ for penguin dominated modes. For all the other modes it can
be neglected.  The initial values of the WC's are derived from the
matching condition at the $m_W$ scale.  However we need to
renormalize them \cite{14,15} when we use these coefficients at
the $m_{b}$ scale. We will use the effective values of WC's at the
scale $\mu =m_b$. It has been shown that in $B\rightarrow PP$ case
there is very little $\mu$ dependence in the final states
\cite{6}.

We obtain the  $c_i(\mu)$'s by solving the following renormalization group
equation:
\begin{eqnarray}
\left( -{\partial \over \partial t} + \beta (\alpha_s) {\partial \over \partial
\alpha_s} \right) {\bf C}(m_W^2/\mu^2, g^2)  = {\hat \gamma^T(g^2)\over 2} {\bf
C}(t, \alpha_s(\mu),\alpha_e),
\end{eqnarray}
where $t\equiv {\rm ln} (M_W^2/\mu^2)$ and  ${\bf C}$ is the
column vector that consists of $(c_i)$'s.   The beta and the gamma
are given by
\begin{eqnarray}
\beta (\alpha_s) &=& - \left(11-{2\over 3}n_f \right){\alpha_s^2\over 16\pi^2}  -
\left(102-{38 \over 3}n_f \right) {\alpha_s^4\over (16\pi^2)^2} + ...\;,
\nonumber\\
\hat \gamma(\alpha_s) &=& \left(\gamma^{(0)}_s
+\gamma^{(1)}_{se}{\alpha_{em}\over{4\pi}}
\right){\alpha_s \over 4\pi}  +\gamma_e^{(0)} {\alpha_{em}\over 4\pi}
+\gamma^{(1)}_s{\alpha^2_s\over(4\pi)^2} +... \;,
\end{eqnarray}   where $\alpha_{em}$ is the electromagnetic coupling and $n_f$
is the number of active quark flavors.

The anomalous-dimension matrices $\gamma^{(0)}_s$ and $\gamma^{(0)}_e$ determine
the leading log  corrections and they are renormalization scheme independent.
The next to leading order corrections  which are determined by
$\gamma^{(1)}_{se}$ and $\gamma^{(1)}_{s}$ are renormalization scheme
dependent.  The $\gamma$'s have been determined in Refs. \cite{14,15}.

We can express $C(\mu)$ (where $\mu$ lies between $M_W$ and $m_b$)
in terms of the initial conditions for the evolution equations
\begin{eqnarray}
C(\mu) = U(\mu,M_W)C(M_W).
\end{eqnarray}
$C(M_W)$'s are obtained from matching the full theory to the
effective theory at the $M_W$ scale \cite{15,16}. The WC's so far
obtained are renormalization scheme dependent. In order to make
them scheme independent we need to use a suitable matrix $T$
\cite{15}. The  WC's at the scale $\mu=m_b$ are given by
\begin{eqnarray}
{\bar C}(\mu) = TU(m_b,M_W)C(M_W).
\end{eqnarray}
The matrix $T$ is given by
\begin{eqnarray}
T ={\bf 1}+ {\hat r}^T_s {\alpha_s\over {4\pi}}+{\hat r}^T_e
{\alpha_e\over{4\pi}},
\end{eqnarray}
where ${\hat r}$ depends on the number of up-type quarks and the down type quarks, respectively. The $r$'s are
given in Ref. \cite{15}.  In order to determine the coefficients at the scale $\mu<m_b$, we need to use the
matching of the evolutions between the scales larger and smaller than the threshold.  In that case, in the
expression for $T$, we need to use $\delta {\hat r}$ instead of ${\hat r}$, where $\delta {\hat
r}=r_{u,d}-r_{u,d-1}$ (where $u$ and $d$ are the number of up type quarks and the number of down type quarks,
respectively). The matrix elements ($O_i's$) are also needed to have one loop correction. The procedure is to
write the one loop matrix element in terms of the tree level matrix element and to generate the effective Wilson
coefficients \cite{17}.
\begin{eqnarray}
\langle c_i O_i \rangle = \sum_{ij} c_i(\mu) \left[\delta_{ij}
+{\alpha_s \over 4\pi} m^s_{ij} +{\alpha_{em}\over 4\pi}m^e_{ij}
\right] \langle O_j \rangle^{{\rm tree}} \;,
\end{eqnarray}
\begin{eqnarray}
\left(\matrix{c^{eff}_1\cr
                               c^{eff}_2\cr
                               c^{eff}_3\cr
                               c^{eff}_4\cr
                               c^{eff}_5\cr
                               c^{eff}_6\cr
                               c^{eff}_7\cr
                               c^{eff}_8\cr
                               c^{eff}_9\cr
                               c^{eff}_{10}}\right) =
                                \left(\matrix{\bar c_1 \cr
                                              \bar c_2 \cr
                                              \bar c_3-P_s/3 \cr
                                              \bar c_4 +P_s \cr
                                              \bar c_5 - P_s/3 \cr
                                              \bar c_6 + P_s \cr
                                              \bar c_7 +P_e \cr
                                              \bar c_8 \cr
                                              \bar c_9 +P_e \cr
                                              \bar c_{10} }\right),
\end{eqnarray}  where
\begin{eqnarray}  P_s &=& {\alpha_s \over 8\pi} \bar c_2 \left[{{V_{cb}
V_{cq}^*}\over {V_{tb} V_{tq}^*}}          \left( {10 \over 9} +G(m_c,\mu,q^2)
\right)  +{{V_{ub} V_{uq}^*} \over {V_{tb} V_{tq}^*}} \left( {10 \over 9}
+G(m_c,\mu,q^2) \right) \right],  \\ \nonumber  P_e &=& {\alpha_{em} \over 9\pi}
(3\bar c_1+\bar c_2) \left[{{V_{cb} V_{cq}^*}
\over {V_{tb} V_{tq}^*}} \left( {10 \over 9} + G(m_c,\mu,q^2) \right)
+{{{V_{ub} V_{uq}^*}\over {V_{tb} V_{tq}^*}}} \left( {10 \over 9} +
G(m_c,\mu,q^2) \right) \right].
\end{eqnarray}  Here $V_{ij}$ are the elements of the CKM matrix.  $m_c$ is the
charm quark mass and $m_u$ is the up quark mass.  The function $G(m,\mu,q^2)$ is
given by
\begin{eqnarray}   G(m,\mu,q^2) = 4\int^1_0 \mbox{d}x x(1-x)
\mbox{ln}{m^2-x(1-x)q^2\over \mu^2} \;,
\end{eqnarray}
where $q$ is the gluon momenta in the penguin diagram
\cite{12A,17,21a}. In the numerical calculation, we will use $q^2
= m_b^2/2$ which represents the average value and the full
expressions for $P_{s,e}$.  In Table I we show the values of the
effective Wilson coefficients at the scale $m_b$ and $m_b/2$ for
the process $b \rightarrow s q \bar q$. Values for $b \rightarrow
d q \bar q$ can be similarly obtained. These coefficients are
scheme independent and gauge invariant.

Since the color octet contribution is neglected in factorization approximation, we keep $\xi \equiv 1/N_c$ as a
variable. As an example, let us consider the $u$ quark contribution of the operator $O_5$ to the process $B^-
\rightarrow \pi^0 K^-$. There are two configurations: (part 1)$= <K^-| \bar s \gamma_{\mu} (1-\gamma_5) b|B^->
<\pi^0| \bar u \gamma_{\mu} (1+\gamma_5) u|0>$. For (part 2), we need to Fierz-transform the operator $O_5$ to $-2
\{ (1/N_c) [\bar u (1-\gamma_5 )b] [\bar s (1+\gamma_5) u)] +(1/2) [\bar u (1-\gamma_5) \lambda_c b] [\bar
s (1+\gamma_5) \lambda_c u]\}$, where $\lambda_c$ is the color matrix. Only the (part 1) contributes in the
factorization approximation. In order to account for this color octet term (part 2), one needs to make $\xi$ a
free parameter.

\section{HADRONIC MATRIX ELEMENTS IN FACTORIZATION APPROXIMATION}
The generalized factorization approximation has been quite
successfully used in two body $D$ decays as well as $B\rightarrow
D$ decays \cite{10}. The method includes color octet
nonfactorizable contribution by treating $\xi \equiv 1/N_c$ as an
adjustable parameter \cite{9,6,7,11b,12A,12a,18}.   In this work
one of our goals is to establish the range of value of a
\emph{single} $\xi$ for the best fit in both $B \rightarrow PP$
and $VP$ decays, where $P$ and $V$ are all \emph{light} mesons
such as $\pi$, $K^{(*)}$, $\rho$, $\omega$, and $\phi$.

Let us describe the parameterizations of the matrix elements and the form factors in the case of $B \rightarrow
PP$ and $VP$ decays.
\begin{eqnarray}
\langle P(p^{\prime})|V_{\mu}|B(p) \rangle    &=& \left[(p^{\prime}+p)_{\mu} -
{m^2_B-m^2_P \over q^2} q_{\mu} \right] F_1(q^2)  +{m^2_B-m^2_P \over q^2}
q_{\mu} F_0(q^2),  \\ \nonumber
\langle V(\epsilon, p^{\prime})|(V_{\mu}-A_{\mu})|B(p)\rangle  &=&{2 \over
{m_B+m_V}}i\epsilon_{\mu\nu\alpha\beta}\epsilon^{\nu*}p^{\alpha}
p^{\prime\beta}V(q^2)  \\ \nonumber  &-& (m_B+m_V)
\left[\epsilon_\mu-{(\epsilon^* \cdot q) \over q^2}q_{\mu} \right] A_1(q^2)  \\
\nonumber &+& {(\epsilon^* \cdot q) \over {m_B+m_V}}
\left[(p+p^{\prime})_{\mu}-{{m_B^2-m_V^2} \over q^2}q_{\mu} \right] A_2(q^2) \\
\nonumber  &-& (\epsilon^* \cdot q) {{2m_V} \over q^2}q_{\mu}A_0(q^2),
\end{eqnarray}  and the decay constants, $f_P$ and $f_V$, are given by
\begin{eqnarray}
\langle 0|A_{\mu}|P(p)\rangle &=& i f_P p_{\mu},  \,\,\,
\langle 0|V_{\mu}|V(\epsilon,p)\rangle =i f_V m_V \epsilon_{\mu},
\end{eqnarray}  where $P$, $V$, $V_{\mu}$, and $A_{\mu}$ denote a pseudoscalar
meson, a vector meson, a vector current, and an axial-vector
current, respectively. $p (p^{\prime})$ and $m_P (m_V)$ are the
momentum of $B$ meson ($P$ or $V$) and the mass of $P (V)$,
respectively.  $q$ is given by $q=p-p^{\prime}$ and
$\epsilon_{\mu}$ is the polarization vector of $V$.    Note that
$F_1(0)=F_0(0)$ and we set $F(q^2 =m_P^2)=F(q^2 =0)$, since these
form factors are assumed to be pole dominated by mesons at scale
$m_B^2$. Among all the form factors in the $\langle V(\epsilon,
p^{\prime})| (V_{\mu}-A_{\mu}) |B(p)\rangle$ matrix element, only
$A_0$ survives when we calculate the full $B\rightarrow VP$ decay
amplitude. The $A_0$ is related to $A_1$ and $A_2$:
\begin{eqnarray}    A_0(0) = {{m_B+m_V}\over {2m_V}}A_1(0)-{{m_B-m_V} \over
{2m_V}}A_2(0).
\end{eqnarray}

For our numerical calculations we use the following values of the decay
constants (in MeV)
\cite{10,7,21a}:
\begin{eqnarray}    f_{\pi} =132,  \,\, f_K =162,  \,\, f_{\rho} =215,  \,\,
f_{\omega} =215,  \,\,  f_{K^*} =225,  \,\, f_{\phi} =237.
\end{eqnarray}  For the values of the form factors, two different sets, based on
the Bauer-Stech-Wirbel (BSW) model \cite{20} and light-cone QCD sum rule
analysis \cite{21},  have been often used in literature \cite{21a}.  In our
analysis, to achieve the best fit,  we treat the form factors as
parameters in a certain range of values which are reasonably consistent with the
frequently used values.  We first start with the following values of form
factors :
$F_0^{B \rightarrow \pi} =0.33$, $F_0^{B \rightarrow K} =0.38$ obtained in the
BSW model,  and $A_0^{B \rightarrow \rho} = A_0^{B \rightarrow \omega} =0.4$.
Then, to find the \emph{single} $\xi$ consistently explaining the experimental
data, we vary the values of the form factors in a certain range as well as other
parameters,  such as the  CKM weak phase $\gamma$ and the strange quark mass
$m_s$.

\section{$B$ DECAYS INTO TWO PSEUDOSCALARS}
Here we consider the decays $B \rightarrow \pi \pi$ and $K \pi$.   The CLEO collaboration has made first
observations of the decay modes $B \rightarrow \pi^+ \pi^-$ and $K^0 \pi^0$, and has presented improved
measurement of BRs for $B  \rightarrow K^{\pm} \pi^{\mp}$ and $B^{\pm} \rightarrow K^0 \pi^{\pm}$ and $B^{\pm}
\rightarrow K^{\pm} \pi^0$ as follows \cite{5} :
\begin{eqnarray}
{\mathcal B} (B^0 \rightarrow \pi^+ \pi^-)  &=& (4.3^{+1.6}_{-1.4}\pm 0.5) \times 10^{-6},
\\ \nonumber  {\mathcal B} (B^0 \rightarrow K^{\pm} \pi^{\mp}) &=& (17.2^{+2.5}_{-2.4}\pm 1.2) \times 10^{-6},
\\ \nonumber  {\mathcal B} (B^{\pm} \rightarrow K^{\pm} \pi^0) &=& (11.6^{+3.0+1.4}_{-2.7-1.3}) \times 10^{-6},
\\ \nonumber  {\mathcal B} (B^{\pm} \rightarrow K^0 \pi^{\pm}) &=& (18.2^{+4.6}_{-4.0}\pm 1.6) \times 10^{-6},
\\ \nonumber  {\mathcal B} (B^0 \rightarrow K^0 \pi^0)  &=& (14.6^{+5.9+2.4}_{-5.1-3.3}) \times 10^{-6}.
\end{eqnarray}

In Figs. 1$-$5, we plot the BRs averaged over particle-antiparticle decays for the modes $\pi^+ \pi^-$, $K^0
\pi^{\pm}$, $K^{\pm} \pi^{\mp}$, $K^{\pm} \pi^0$, and $K^0 \pi^0$ in final states as a function of $\xi = 1/N_c$
for $\mu =m_b$.  In these figures, we use five different kinds of lines corresponding to different values of
parameters, in addition to thick solid lines representing the experimental upper and lower bounds.  The solid line
corresponds to the case of choosing the form factors $F_0^{B \rightarrow \pi}(0) = 0.33$ and $F_0^{B \rightarrow
K}(0) = 0.38$ based on the BSW model,
 $\gamma \equiv {\rm Arg}(V_{ub}^*) =60^0$, $m_s (m_b) =106$ MeV, $V_{cb}
=0.040$,
$|V_{ub} /V_{cb}| =0.087$, and $|V_{td}| =0.004$.  The short dashed line then
corresponds to the case of choosing $\gamma =110^0$ and the other parameters are
the same as  in the solid line case.  Similarly, the dot-dash-dot line corresponds
to the case of
 choosing $m_s (m_b) =85$ MeV, while the values of other parameters are the same
as   in the solid line case.  For the dot-dashed and long dashed lines, we
choose $m_s (m_b) =106$ MeV and 85 MeV, respectively, with smaller values of
form factors $F_0^{B \rightarrow \pi}(0) = 0.26$ and $F_0^{B
\rightarrow K}(0) = 0.29$, and $\gamma =110^0$.   Thus, by comparing each line
with the solid line, one can easily see how the decay rate for any mode changes
as a particular parameter, such as $\gamma$ or $m_s$, changes.  Note that in
Fig. 1 for $B^0 \rightarrow
\pi^+ \pi^-$ decay, the dot-dash-dot and long dashed lines are identical to the
solid and  dot-dashed lines, respectively, since the amplitude for this decay
mode does not depend on $m_s$.   Similarly, in Fig. 2 for $B^{\pm} \rightarrow
K^0 \pi^{\pm}$ decay, the short dashed line is  identical to the solid line
since the amplitude for this mode receives contribution from the  penguin
diagram only and is independent of $\gamma$.

The decay rate for $B^0 \rightarrow \pi^+ \pi^-$ is proportional
to $|F_0^{B \rightarrow \pi}|^2$ and is sensitive to the value of
the form factor.   In Fig. 1, one can see that the BR decreases,
as the value of $F_0^{B \rightarrow \pi}$ decreases and/or the
value of $\gamma$ increases.  In order to fit the experimental
upper limit on the BR for this mode, a smaller $F_0^{B \rightarrow
\pi}$ and a larger $\gamma$ (dot-dashed line, identical to the
long dashed line) are favored.   For $F_0^{B \rightarrow \pi}(0) =
0.26$ and $\gamma =110^0$, the values of $\xi \gtrsim 0.44$ are
allowed.  The rate for $B^{\pm} \rightarrow K^0 \pi^{\pm}$ is also
proportional to $|F_0^{B \rightarrow \pi}|^2$ and is sensitive to
$m_s$ as well.  Figure 2 shows that the BR increases, as the value
of $F_0^{B \rightarrow \pi}$ increases and/or the value of $m_s$
decreases.   In order to find a solution consistent with $B^0
\rightarrow \pi^+ \pi^-$, $F_0^{B \rightarrow \pi}$ should not be
too small and smaller $m_s$ is favored (long dashed line).    For
$F_0^{B \rightarrow \pi}(0) = 0.26$ and $m_s (m_b) =85$ MeV (long
dashed line), the allowed values of $\xi$ are $\xi \lesssim 0.45$.
However, for larger $m_s (m_b) =106$ MeV (dot-dashed line),  the
allowed values of $\xi$ are smaller $\xi \lesssim 0.15$ and these
values of $\xi$ are not  allowed by the dot-dashed line for $B^0
\rightarrow \pi^+ \pi^-$.  In Fig. 3, we plot the BR for $B^0
\rightarrow K^{\pm} \pi^{\mp}$ averaged over particle-antiparticle
decays as a function of $\xi$ for $\mu =m_b$. This decay mode is
sensitive to $F_0^{B \rightarrow \pi}$, $\gamma$ and $m_s$. In the
case of the long dashed line, the allowed values of $\xi$ are $\xi
\lesssim 0.6$,  which are consistent with those in $B^0
\rightarrow \pi^+ \pi^-$ and $B^+ \rightarrow K^0 \pi^+$.
Similarly, in Figs. 4 and 5, we plot the BRs for $K^{\pm} \pi^0$
and $K^0 \pi^0$  in the final states as a function of $\xi$ for
$\mu =m_b$. These modes depend on $F_0^{B \rightarrow \pi}$,
$F_0^{B \rightarrow K}$, $m_s$, and $\gamma$. In the long dashed
line case, $\xi \lesssim 0.88$ and $\xi \lesssim 0.47$ are allowed
for $B \rightarrow K^{\pm} \pi^0$ and $K^0 \pi^0$, respectively,
and are consistent with those in the above decays (Figs. 1$-$3).
Therefore, we conclude that the long dashed lines  (the dot-dashed
line in the case of $B \rightarrow \pi^+ \pi^-$) in Figs. 1$-$5
represent the possible solution compatible with all the
experimental limits on the BRs for decay modes $\pi^+ \pi^-$, $K^0
\pi^{\pm}$, $K^{\pm}\pi^{\mp}$, $K^{\pm} \pi^0$, and $K^0 \pi^0$
in the final states.  The values of $\xi$ allowed by the data are
those near $\xi \approx 0.45$. (In fact, we shall see that the
long dashed line can also consistently explain all  the
experimental data for $B \rightarrow VP$ considered in next
section.)   Note that the modes $B \rightarrow \pi^+ \pi^-$, $K^0
\pi^{\pm}$, and $K^0 \pi^0$ provide strong constraints on the
values of $\xi$ to satisfy the data.  In particular, the data on
the modes $B \rightarrow \pi^+ \pi^-$ and $K^0 \pi^{\pm}$ put very
tight limits on the allowed values of $\xi$. Hence, an improved
measurement of these modes will be important in testing the
framework of factorization.

The experimental bounds on the BRs for decays $B \rightarrow PP$ provide constraints on the parameters.  The
favored values of the parameters are
\begin{eqnarray}
& \mbox{}& F_0^{B \rightarrow \pi}(0) = 0.26, \;\;\;   F_0^{B \rightarrow K}(0) = 0.29,  \\ \nonumber
& \mbox{}&
\gamma \approx 110^0, \hspace{1cm} m_s (m_b) =85 \; {\rm MeV}, \\ \nonumber
& \mbox{}& V_{cb} =0.040, \;\;\;
|V_{ub} /V_{cb}| =0.087, \;\;\;  |V_{td}| =0.004.
\end{eqnarray}
Current best estimates for CKM matrix elements are $V_{cb} =0.0381 \pm 0.0021$ and $|V_{ub} /V_{cb}| =0.085 \pm
0.019$ \cite{22}.   The CLEO has recently made the first determination of the value of $\gamma =113^{0+25^0}_{\;\;
-23^0}$ by any method other than the unitarity triangle construction \cite{5,5a,5aa}.  The favored values for the
CKM matrix elements in our analysis above are chosen to get the best fit for the experimental limits on the BRs
for the decay processes $B \rightarrow PP$ and $VP$.   We find that, if $|V_{ub}|$ increases, the rates for $\pi^+
\pi^-$ and $\omega h$ $(h =\pi, K)$ increase, while the rate for $K^{\pm} \pi^0$ decreases.  Also if $|V_{td}|$
increases, then the rates for $\pi^{\pm} \pi^{\mp}$ and $K^{\pm} \pi^0$ increase and the rate for $\omega h$
decreases.

In Table II, we present the BRs and the CP asymmetries for $B \rightarrow PP$ decays at a representative value of
$\xi =0.45$. (We shall see that the values of $\xi$ near $\xi \approx 0.45$ are favored to fit all the data.)
Available experimental values are also presented. The BRs for all the modes are compatible with the present
experimental data.  The CP asymmetry, ${\mathcal A}_{CP}$, is defined by
\begin{eqnarray}
{\mathcal A}_{CP} = {{\mathcal B}(b \rightarrow f)  -{\mathcal
B}(\bar b \rightarrow \bar f) \over {\mathcal B}(b \rightarrow f)  +{\mathcal
B}(\bar b \rightarrow \bar f)},
\end{eqnarray}
where $b$ and $f$ denote $b$ quark and a generic final state, respectively.   The recent CLEO search for CP
asymmetries in $B \rightarrow K \pi$ decays has found : $-0.70 \leq {\mathcal A}_{CP} (B^{\pm} \rightarrow K^{\pm}
\pi^0) \leq 0.16$, $-0.35 \leq {\mathcal A}_{CP} (B^0 \rightarrow K^{\pm} \pi^{\mp}) \leq 0.27$ at 90$\%$
confidence level (C.L.).  The expected CP asymmetries in $B \rightarrow PP$ decays are generally small and range
from $-11 \%$ to 0.

The BRs in this analysis have been evaluated at the scale $m_b$. However, if we had chosen the QCD scale
$\mu=m_b/2$, the result would not change much \cite{6}. We also see that the favored values of parameters involve
a lighter strange quark mass. This, however, is in accordance with the latest trend of lattice results \cite{lat}.
The ratios of the quark masses are much better known than the individual masses. For example \cite{lr},
\begin{eqnarray}
{m_u \over m_d} = 0.553 \pm 0.043;\,\,\,\, {m_s \over m_d}=18.9 \pm 0.8
\label{qr}
\end{eqnarray}
The strange quark mass $m_s$ is in considerable doubt: i.e., QCD
sum rules give $m_s(1\; {\rm GeV})=(175\pm 25)$ MeV and lattice
gauge theory gives $m_s(2\; {\rm GeV})=(100 \pm 20 \pm 10)$ MeV in
the quenched lattice calculation \cite{lat}. In this analysis we
have varied $m_s$ from 150 to 116 MeV at 1 GeV scale. We see that
the $m_s$ of 116 MeV gives rise to the best fit. We have used the
quark masses at the $m_b$ scale. The magnitude of $m_s$ reduces
from 150 to 106 MeV and from 116 to 85 MeV at the $m_b$ scale
through 3 loop QCD and 1 loop QED RGEs. The magnitudes of the
other quark masses ($m_u$ and $m_d$) also depend on the strange
quark mass. Satisfying the constraints from Eq. (\ref{qr}), the
values of $m_d$ we have used are 5.9 MeV (corresponding to
$m_s$=150 MeV) and 4.7 MeV (corresponding to $m_s$=116 MeV) at the
$m_b$ scale.  Similarly, the values of $m_u$ we have used are 3.4
MeV (corresponding to $m_s$=150 MeV) and 2.8 MeV (corresponding to
$m_s$=116 MeV) at the $m_b$ scale.

The difference in $m_s$ affects the BRs for the $B \rightarrow \pi K$ modes, since the amplitudes contain a factor
such as $X = M_K^2 /((m_b + m_u)(m_s + m_u))$.  On the other hand, changes in $m_d$ or $m_u$ do not affect the BRs
significantly. In the case of $|\Delta S|=1$ decays, $m_u$ or $m_d$ can always be neglected compared to $m_b$ or
$m_s$ in the factor $X$. In the case of $\Delta S=0$ decays, the tree contribution is large compared to the
penguin contribution and the factor similar to $X$ appears only in the penguin term. Hence the effect is small.
For example, the BR for $B^0 \rightarrow \pi^+ \pi^-$ changes from $6.00 \times 10^{-6}$ to $6.10 \times 10^{-6}$
at $\xi=0.45$, when one changes the $m_d$ from 5.9 MeV to 4.7 MeV (changing the $m_u$ mass as well). Similar
results hold true also in the $B \rightarrow VP$ case.

\section{$B$ DECAYS INTO A VECTOR AND A PSEUDOSCALAR}
We now analyze the decay processes $B \rightarrow VP$ which include $B \rightarrow \omega \pi (K)$, $\rho \pi
(K)$, $\phi \pi (K)$, and $K^* \pi (K)$. The recent measurement at CLEO has yielded the following bounds \cite{2}
:
\begin{eqnarray}
\label{rhopi} {\mathcal B} (B^{\pm} \rightarrow \omega \pi^{\pm})  &=& (11.3^{+3.3}_{-2.9} \pm 1.4) \times
10^{-6},
\\ \nonumber {\mathcal B} (B^{\pm} \rightarrow \omega h^{\pm})  &=& (14.3^{+3.6}_{-3.2} \pm 2.0) \times 10^{-6},
\\ \nonumber {\mathcal B} (B^{\pm} \rightarrow \rho^0 \pi^{\pm})  &=& (10.4^{+3.3}_{-3.4} \pm 2.1) \times 10^{-6},
\\ \nonumber {\mathcal B} (B^0 \rightarrow \rho^{\pm} \pi^{\mp})  &=& (27.6^{+8.4}_{-7.4}\pm 4.2) \times 10^{-6},
\\ \nonumber {\mathcal B} (B^0 \rightarrow K^{*\pm} \pi^{\mp})  &=& (22^{+8+4}_{-6-5}) \times 10^{-6},
\end{eqnarray}
and
\begin{eqnarray}
{\mathcal B} (B^{\pm} \rightarrow \omega K^{\pm}) &<& 7.9 \times 10^{-6}, \label{omegaK}
\\ \nonumber {\mathcal B} (B^{\pm} \rightarrow \phi K^{\pm}) &=& (6.4^{+2.5+0.5}_{-2.1-2.0}) \times 10^{-6},
\end{eqnarray}
where $h^{\pm}$ denotes $\pi^{\pm}$ or $K^{\pm}$, and the BR for $B^0 \rightarrow \rho^{\pm} \pi^{\mp}$ is the sum
of the BRs for $B^0 \rightarrow \rho^+ \pi^-$ and $B^0 \rightarrow \rho^- \pi^+$.  Note that the above BR for $B^0
\rightarrow K^{*\pm} \pi^{\mp}$ still involves a large error.

As in the case of $B \rightarrow PP$ decays, in Figs. 6$-$11, we plot the BRs averaged over particle-antiparticle
decays for the modes $B \rightarrow \omega \pi^{\pm}$, $\omega h^{\pm}$, $\omega K^{\pm}$, $\rho^{\pm} \pi^{\mp}$,
$\phi K^{\pm}$, and $K^{*\pm} \pi^{\mp}$ as a function of $\xi$ for $\mu =m_b$. Six different kinds of lines are
used, corresponding to different values of parameters.   The definitions of the (five) lines are the same as those
in $B \rightarrow PP$ case, except that the form factors $A_0(0) \equiv A_0^{B \rightarrow \rho}(0) =A_0^{B
\rightarrow \omega}(0) =0.4$ are now added.  The dotted line is newly introduced, which corresponds to $A_0(0)
=0.36$ with the same values of other parameters as those in the solid line case.    Thus, a comparison of the
dotted line with the solid line shows how the BR for a particular mode changes as $A_0$ changes.

In Fig. 6, we present the plot of the BR for $B^{\pm} \rightarrow \omega \pi^{\pm}$ as a function of $\xi$.  This
decay mode receives the dominant contribution from the tree diagram and is sensitive to the form factors $A_0^{B
\rightarrow \omega}$, $F_1^{B \rightarrow \pi}$, and the weak phase $\gamma$. The long dashed line and the
dot-dash-dot line are identical to the dot-dashed line and the solid line, respectively, since the rate for this
mode does not depend on $m_s$.  All the lines are well within the experimental limits for values of $\xi$ in a
broad region.   In the case of the long dashed line, the allowed values of $\xi$ are $0.3 \lesssim \xi \lesssim
0.75$.    The recent CLEO search for CP asymmetry in $B^{\pm} \rightarrow \omega \pi^{\pm}$ decay has found :
$-0.80 \leq {\mathcal A}_{CP} (B^{\pm} \rightarrow \omega \pi^{\pm}) \leq 0.12$ at 90$\%$ C.L.   We find that the
expected CP asymmetry in this mode is $-11\%$ for a representative value of $\xi =0.45$ (we shall see below that
the values of $\xi$ near $\xi \approx 0.45$ are the favored values for the best fit).

The plot of the BR for $B^{\pm} \rightarrow \omega K^{\pm}$ as a function of $\xi$ is shown in Fig. 7.  The rate
for this process depends on $A_0^{B \rightarrow \omega}$, $F_1^{B \rightarrow K}$, $\gamma$, and $m_s$.   The
previous experimental result from CLEO for this decay mode \cite{23} showed the large BR of ${\mathcal B} (B^{\pm}
\rightarrow \omega K^{\pm}) =(15^{+7}_{-6} \pm 2) \times 10^{-6}$, but in the recent CLEO report \cite{2} the
statistical significance for this mode is only $2.1 \sigma$ and the upper limit for the BR at 90$\%$ C.L. has been
set as in Eq. (\ref{omegaK}), which is much lower than the previous one. Thus, as can be seen in Fig. 7, the
values of $\xi$ in a broad region are compatible with the experimental upper limit. However, in our earlier work
\cite{7}, only smaller values of $\xi \lesssim 0.05$ or larger values of $\xi
> 0.6$ were allowed to fit the previous data.  The allowed values of $\xi$ for the long dashed line are $\xi
\lesssim 0.67$.  At a representative value of $\xi =0.45$, the expected BR for this mode is ${\mathcal B} (B^{\pm}
\rightarrow \omega K^{\pm}) = 1.33 \times 10^{-6}$.

Figure 8 shows the plot of the BR for $B^{\pm} \rightarrow \omega
h^{\pm}$ as a function of $\xi$, where $h$ is $\pi$ or $K$.  To
obtain the best fit for this mode, smaller values of $F_1^{B
\rightarrow \pi}$, $F_1^{B \rightarrow K}$, and $\gamma$, and
larger values of $A_0^{B \rightarrow \omega}$ and $m_s$ are
favored.  The previous CLEO measurement of the BR for this mode
\cite{23} was ${\mathcal B} (B^{\pm} \rightarrow \omega h^{\pm})
=(25^{+8}_{-7} \pm 3) \times 10^{-6}$, but the recent value
\cite{2} has been reduced to ${\mathcal B} (B^{\pm} \rightarrow
\omega h^{\pm}) =(14.3^{+3.6}_{-3.2} \pm 2.0) \times 10^{-6}$.
Thus, the values of $\xi \lesssim 0.1$ and $0.4 \lesssim \xi
\lesssim 0.64$ for the long dashed line are compatible with the
recent data in this mode, while the previous allowed values of
$\xi$ are $\xi \approx 0$ and $\xi \gtrsim 0.5$.    The expected
BR for this mode at a representative value of $\xi =0.45$ is
${\mathcal B} (B^{\pm} \rightarrow \omega h^{\pm}) = 11.34 \times
10^{-6}$.

The case of $B^{\pm} \rightarrow \phi K^{\pm}$ is shown in Fig. 9.  The previous CLEO result for this mode
\cite{23} was ${\mathcal B} (B^{\pm} \rightarrow \phi K^{\pm}) < 5 \times 10^{-6}$ at 90 $\%$ C.L. but recently
CLEO has announced the new data \cite{2} : ${\mathcal B} (B^{\pm} \rightarrow \phi K^{\pm}) =
(6.4^{+2.5+0.5}_{-2.1-2.0}) \times 10^{-6}$, ${\mathcal B} (B^0 \rightarrow \phi K^0_S) < 12 \times 10^{-6}$, and
the combined branching ratio ${\mathcal B} (B \rightarrow \phi K) = (6.2^{+2.0+0.7}_{-1.8-1.7}) \times 10^{-6}$.
(The Belle Collaboration has recently reported the branching ratio \cite{2} : ${\mathcal B} (B^{\pm} \rightarrow
\phi K^{\pm}) = (17.2^{+6.7}_{-5.4} \pm 1.8) \times 10^{-6}$, which is very large and inconsistent with the CLEO
data. To be consistent, in this analysis we use the recent CLEO data only. Future improved data for this mode are
called for.) This decay is a pure penguin process and is sensitive to $F_1^{B \rightarrow K}$, but independent of
$A_0$, $\gamma$, and $m_s$. A smaller $F_1^{B \rightarrow K}$ is favored for a better fit in this decay.  But the
decays $B \rightarrow K \pi$ disfavor too small values of $F_{0,1}^{B \rightarrow \pi}$ and $F_{0,1}^{B
\rightarrow K}$. We find that $F_0^{B \rightarrow \pi}(0) \approx 0.26$, $F_0^{B \rightarrow K}(0) \approx 0.29$
are favored. For the dot-dashed line, identical to the long dashed line in this mode, the values of $0.3 \lesssim
\xi \lesssim 0.55$ are compatible with the data.

In Fig. 10, we present the cases of $B \rightarrow \rho \pi$ decays in which the tree contribution is dominant. In
(a), the sum of the BRs for $B^0 \rightarrow \rho^+ \pi^-$ and $B^0 \rightarrow \rho^- \pi^+$ as a function of
$\xi$ is shown in order to compare with the recent CLEO data in Eq. (\ref{rhopi}).  The decay $B^0 \rightarrow
\rho^+ \pi^-$ is sensitive to $F_1^{B \rightarrow \pi}$ and $\gamma$, while $B^0 \rightarrow \rho^- \pi^+$ is
sensitive to $A_0^{B \rightarrow \rho}$ and $\gamma$.  The lines are well within the experimental upper and lower
limits for most values of $\xi$.  For the dot-dashed line, identical to the long dashed line, the allowed values
of $\xi$ are $\xi \gtrsim 0.08$.    The expected BR for this mode is ${\mathcal B} (B^0 \rightarrow \rho^{\pm}
\pi^{\mp}) = 29.41 \times 10^{-6}$ at a representative value of $\xi =0.45$.   The case of $B^{\pm} \rightarrow
\rho^0 \pi^{\pm}$ is shown in (b).   This process is sensitive to $A_0^{B \rightarrow \rho}$, $F_1^{B \rightarrow
\pi}$, and $\gamma$.   For the dot-dashed line, identical to the long dashed line, the values of $\xi \gtrsim
0.34$ are compatible with the experimental data.  This can be compared with the case of $B^{\pm} \rightarrow
\omega \pi^{\pm}$ shown in Fig. 6.  These two modes are both tree-dominated and have similar values of masses,
decay constants, and form factors.   So it is expected that their BRs are not very different : the recent CLEO
data for $\rho^0 \pi^{\pm}$ and $\omega \pi^{\pm}$ decays are $(10.4^{+3.3}_{-3.4} \pm 2.1) \times 10^{-6}$ and
$(11.3^{+3.3}_{-2.9} \pm 1.4) \times 10^{-6}$, respectively, as in Eq. (\ref{rhopi}). The favored values in our
analysis are ${\mathcal B} (B^{\pm} \rightarrow \rho^0 \pi^{\pm}) = 10.06 \times 10^{-6}$ and ${\mathcal B}
(B^{\pm} \rightarrow \omega \pi^{\pm}) = 10.01 \times 10^{-6}$ at a representative value of $\xi =0.45$, which are
consistent with the recent data.

Figure 11 shows the plot of ${\mathcal B} (B^0 \rightarrow
K^{*\pm} \pi^{\mp})$ as a function of $\xi$.  This mode receives
the dominant contribution from the penguin diagram.    Our
theoretical expectation of the BR for this decay is less than the
experimental limits at 1$\sigma$ level (thick lines), but is
greater than the lower limit at 2$\sigma$ level (gray line).
Thus, within 2$\sigma$ range, our result is compatible with the
data for this mode.   Since the measurement of this decay still
involves large error, an improvement in the experiment will be
crucial to test the framework of our work.

In Tables III and IV, we present the BRs and the CP asymmetries
for the decays $B \rightarrow VP$ ($\Delta S =0$ and $|\Delta S|
=1$) at a representative value of $\xi =0.45$.  Available
experimental results are also presented. All the theoretical
values are compatible with the present experimental bounds.  In
particular, in some decay modes, the CP asymmetries are expected
to be large. Among $\Delta S =0$ decays, the CP asymmetry is
expected to be relatively large in a few decay modes: (i) in $B^0
\rightarrow \rho^0 \pi^0$, the expected CP asymmetry is $-27\%$
with the expected BR of $0.65 \times 10^{-6}$, (ii) in $B^0
\rightarrow \omega \pi^0$, the CP asymmetry is expected to be
$-18\%$ with the expected BR of $0.032 \times 10^{-6}$, (iii) in
$B^{\pm} \rightarrow \omega \pi^{\pm}$, the CP asymmetry is
expected to be $-11\%$ with the expected BR of $10.01 \times
10^{-6}$. Among $|\Delta S| =1$ decays, there are several
interesting modes: (i) the expected CP asymmetry in $B^0
\rightarrow \omega K^0$ is $-29\%$ with the expected BR of $0.15
\times 10^{-6}$, (ii) the expected CP asymmetry in $B^{\pm}
\rightarrow \omega K^{\pm}$ is $-19\%$ with the expected BR of
$1.33 \times 10^{-6}$, (iii) the CP asymmetry in $B^0 \rightarrow
K^{*\pm} \pi^{\mp}$ is expected to be $-14 \%$ with the expected
BR of $6.27 \times 10^{-6}$, (iv) the CP asymmetry in $B^{\pm}
\rightarrow K^{*\pm} \pi^0$ is expected to be $15\%$ with the
expected BR of $3.56 \times 10^{-6}$.

\section{CONCLUSION}
Motivated by the recent CLEO data, we have analyzed charmless hadronic two body decays of $B$ mesons $B
\rightarrow PP$ and $B \rightarrow VP$.   In the framework of generalized factorization,  we have carefully
examined the values of several input parameters  to which the predictions are sensitive. Those input parameters
are the form factors, the strange quark mass, $\xi \equiv 1/ N_c$, the CKM matrix elements, and in particular, the
weak phase $\gamma$.

We have found that the experimental bounds on the BRs for the decay modes $B \rightarrow \pi^+ \pi^-$, $K^0
\pi^{\pm}$, and $K^0 \pi^0$ among $B \rightarrow PP$ modes put strong constraints on the parameters.  The
constraints on parameters from the decays $B^{\pm} \rightarrow \omega h^{\pm}$ among $B \rightarrow VP$ are also
strong and lead to the following favored values of the parameters for the best fit (in Figs. 1$-$11, the long
dashed line (or the dot-dashed line when the long dashed line is absent) represents the case corresponding to the
best fit):
\begin{eqnarray}
&\mbox{}& \xi \approx 0.45                                                         \\ \nonumber &\mbox{}& F_0^{B
\rightarrow \pi}(0) = 0.26, \;\;\; F_0^{B \rightarrow K}(0) = 0.29,             \\ \nonumber &\mbox{}&A_0^{B
\rightarrow \rho}(0) = 0.4, \;\;\; A_0^{B \rightarrow \omega}(0) = 0.4,          \\ \nonumber &\mbox{}& \gamma
\approx 110^0, \hspace{1cm}  m_s (m_b) =85 \; {\rm MeV},                        \\ \nonumber &\mbox{}& V_{cb}
=0.040, \;\;\; |V_{ub} /V_{cb}| =0.087, \;\;\;   |V_{td}| =0.004.
\end{eqnarray}
[In fact, a little smaller values of $A_0(0)$ are also allowed.]
It has been known that there exists the discrepancy in values of
$\gamma$ extracted from the CKM-fitting at $\rho - \eta$
plan\cite{final} and from the $\chi^2$ analysis of hadronic decays
of $B$ mesons\cite{5aa}. The value of $\gamma$ obtained from the
each case is $\gamma = 60^0 \sim 80^0$ from the CKM-fitting at
$\rho - \eta$ plane, or $\gamma = 90^0 \sim 140^0$ from the
hadronic $B$ decay analysis. In our analysis we find that $\gamma
\approx 110^0$ is favored to fit the given data. We have shown
that the recent CLEO data in $B \rightarrow PP$ and $VP$ modes can
be satisfactorily explained with $\xi \approx 0.45$, except for
the BR of the decay mode $B^0 \rightarrow K^{*\pm} \pi^{\mp}$ at
1$\sigma$ level [at 2$\sigma$ level, our prediction for ${\mathcal
B} (B^0 \rightarrow K^{*\pm} \pi^{\mp})$ is compatible with the
data].   An improved measurement of the BR for this process will
be crucial in testing the framework of factorization.    We have
also identified the decay modes where the CP asymmetries are
expected to be large,  such as $B \rightarrow \rho^0 \pi^0$,
$\omega \pi^0$, $\omega \pi^{\pm}$ in $\Delta S =0$ decays, and $B
\rightarrow \omega K^0$, $\omega K^{\pm}$, $K^{*\pm} \pi^0$,
$\rho^0 K^0$ in $|\Delta S| =1$ decays.
\\

\centerline{\bf ACKNOWLEDGEMENTS}
\medskip

\noindent We would like to thank J. G. Smith for useful
conversations and comment. This work was supported in part by
National Science Foundation Grant No. PHY-9722090  and by the US
Department of Energy Grant No. DE FG03-95ER40894.

\newpage
\begin{table}
\caption{Effective Wilson coefficients for the $b \rightarrow s$ transition at the scales $\mu =m_b$ and $m_b/2$.}
\vspace{0.5cm}
\begin{tabular}{ccc}
  WC's & $\mu =m_b$ & $\mu =m_b/2$ \\ \hline
$c^{\rm eff}_1$ & 1.149 & 1.135  \\ $c^{\rm eff}_2$ & $-0.3209$ & $-0.282$  \\ $c^{\rm eff}_3$ & $0.02175 -0.00414
i$ & $0.0228718 +0.004689 i$  \\ $c^{\rm eff}_4$ & $-0.04906 -0.01242 i$ & $-0.051144 -0.004689 i$  \\ $c^{\rm
eff}_5$ & $0.01560 +0.00414 i$ & $-0.051144 -0.004689 i$  \\ $c^{\rm eff}_6$ & $-0.06063 -0.01242 i$ & $-0.0653549
-0.0140673 i$  \\ $c^{\rm eff}_7$ & $-0.000859 +0.000073 i$ & $0.00122773 +0.00005724 i$  \\ $c^{\rm eff}_8$ &
0.001433 & $-0.0000953211$  \\ $c^{\rm eff}_9$ & $-0.011487 +0.000073 i$ & $-0.0120155 +0.0000572433 i$  \\
$c^{\rm eff}_{10}$ & 0.003174 & 0.00218628
\end{tabular}
\end{table}

\begin{table}
\caption{The branching ratios $({\mathcal B})$ and the CP asymmetries $({\mathcal A}_{CP})$ for $B$ decay modes
into two pseudoscalar mesons at a representative value of $\xi =0.45$.  Charged conjugate modes are implied.  The
experimental ${\mathcal A}_{CP}$'s represent 90$\%$ confidence level intervals. } \vspace{0.5cm}
\begin{tabular}{ccccc}  Decay modes & ${\mathcal B}$ $(10^{-6})$ & Experimental
${\mathcal B}$ $(10^{-6})$ &  ${\mathcal A}_{CP}$ & Experimental ${\mathcal A}_{CP}$
\\ \hline $B^+ \rightarrow \pi^+ \pi^0$ & 4.38 & $< 12.7$ & 0 &
\\ $B^0 \rightarrow \pi^+ \pi^-$ & 6.00 & $4.3^{+1.6}_{-1.4}\pm 0.5$ & $-0.11$ &
\\ $B^0 \rightarrow K^0 K^+$ & 0.11 & $< 5.7$ & 0 &
\\ \\ $B^+ \rightarrow K^+ \pi^0$ & 10.14 & $11.6^{+3.0+1.4}_{-2.7-1.3}$ & $-0.061$ & $[-0.70, 0.16]$
\\ $B^+ \rightarrow K^0 \pi^+$ & 13.34 & $18.2^{+4.6}_{-4.0}\pm 1.6$ & 0 &
\\ $B^0 \rightarrow K^+ \pi^-$ & 15.90 & $17.2^{+2.5}_{-2.4}\pm 1.2$ & $-0.066$ & $[-0.35, 0.27]$
\\ $B^0 \rightarrow K^0 \pi^0$ & 8.70 & $14.6^{+5.9+2.4}_{-5.1-3.3}$ & $-0.01$ &
\end{tabular}
\end{table}

\begin{table}
\caption{The branching ratios $({\mathcal B})$ and the CP asymmetries $({\mathcal A}_{CP})$ for $B$ decay modes
($\Delta S =0$) into a vector and a pseudoscalar meson at a representative value of $\xi =0.45$.  Charged
conjugate modes are implied. The experimental branching ratio with $^{\dagger}$ below is the sum of $B^0
\rightarrow \rho^+ \pi^-$ and $\rho^- \pi^+$.  The experimental ${\mathcal A}_{CP}$ represents a 90$\%$ confidence
level interval.} \vspace{0.5cm}
\begin{tabular}{ccccc}  Decay modes & ${\mathcal B}$ $(10^{-6})$ & Experimental
${\mathcal B}$ $(10^{-6})$ & ${\mathcal A}_{CP}$ & Experimental ${\mathcal A}_{CP}$
\\ \hline
$B^+ \rightarrow \omega \pi^+$ & 10.01 & $11.3^{+3.3}_{-2.9}\pm 1.4$ & $-0.11$ & $[-0.80, 0.12]$
\\ $B^+ \rightarrow \rho^0 \pi^+$ & 10.06 & $10.4^{+3.3}_{-3.4} \pm 2.1$ & 0.047
\\ $B^+ \rightarrow \rho^+ \pi^0$ & 10.51 & $< 43$ & $-0.039$ &
\\ $B^+ \rightarrow \bar K^{*0} K^+$ & 0.043 & $< 5.3$ & 0 &
\\ $B^+ \rightarrow K^{*+} \bar K^0$ & 0.013 &  & 0 &
\\ $B^+ \rightarrow \phi \pi^+$ & 0.0028 & $< 4.0$ & 0 &
\\ \\ $B^0 \rightarrow \omega \pi^0$ & 0.032 & $< 5.5$ & $-0.18$ &
\\ $B^0 \rightarrow \rho^+ \pi^-$ & 15.39 & $27.6^{+8.4}_{-7.4} \pm 4.2^{\dagger}$ & $-0.072$ &
\\ $B^0 \rightarrow \rho^- \pi^+$ & 14.02 &  & $-0.015$ &
\\ $B^0 \rightarrow \rho^0 \pi^0$ & 0.65 & $< 5.5$ & $-0.27$ &
\\ $B^0 \rightarrow K^{*0} \bar K^0$ & 0.13 &  & 0 &
\\ $B^0 \rightarrow \phi \pi^0$ & 0.0014 & $< 5.4$ & 0 &
\end{tabular}
\end{table}

\begin{table}
\caption{The branching ratios $({\mathcal B})$ and the CP asymmetries $({\mathcal A}_{CP})$ for $B$ decay modes
($|\Delta S| =1$) into a vector and a pseudoscalar meson at a representative value of $\xi =0.45$.  Charged
conjugate modes are implied.  The experimental branching ratio with $^{*}$ below is from Ref. [34].}
\vspace{0.5cm}
\begin{tabular}{cccc}  Decay modes & ${\mathcal B}$ $(10^{-6})$ & Experimental
${\mathcal B}$ $(10^{-6})$ &  ${\mathcal A}_{CP}$
\\ \hline $B^+ \rightarrow \omega K^+$ & 1.33 & $< 7.9$ & $-0.19$
\\ $B^+ \rightarrow \rho^0 K^+$ & 0.67 & $< 17$ & 0.13
\\ $B^+ \rightarrow \rho^- K^0$ & 1.20 & $< 48^{*}$ & 0
\\ $B^+ \rightarrow K^{*0} \pi^+$ & 4.06 & $< 16$ & 0
\\ $B^+ \rightarrow K^{*+} \pi^0$ & 3.56 & $< 31$ & 0.15
\\ $B^+ \rightarrow \phi K^+$ & 6.56 & $6.4^{+2.5+0.5}_{-2.1-2.0}$ & 0
\\ \\ $B^0 \rightarrow \omega K^0$ & 0.15  & $< 21$ & $-0.29$
\\ $B^0 \rightarrow \rho^- K^+$ & 1.28 & $< 32$ & 0.10
\\ $B^0 \rightarrow \rho^0 K^0$ & 0.082 & $< 27$ &0.17
\\ $B^0 \rightarrow K^{*0} \pi^0$ &0.76  & $< 3.6$ & $-0.13$
\\ $B^0 \rightarrow K^{*+} \pi^-$ & 6.27 & $22^{+8+4}_{-6-5}$ & $-0.14$
\\ $B^0 \rightarrow \phi K^0$ & 6.56 & $< 12$ & 0
\end{tabular}
\end{table}

\newpage
\begin{figure}[htb]
\vspace{1 cm}

\centerline{ \DESepsf(dobphypi1pi1new1.epsf width 12 cm) }
\smallskip
\caption {Branching ratio for $B^0 \rightarrow \pi^+ \pi^-$ as a function of
$\xi(\equiv {1\over N_c})$.
\newline The solid line :
$F_0^{B\rightarrow\pi}(0)$=0.33, $F_0^{B\rightarrow K}(0)$=0.38, $\gamma=60^0$
and $m_s(m_b)$=106 MeV.
\newline The short dashed line :
$F_0^{B\rightarrow\pi}(0)$=0.33, $F_0^{B\rightarrow K}(0)$=0.38, $\gamma=110^0$
and $m_s(m_b)$=106 MeV.
\newline The dot-dashed line :
$F_0^{B\rightarrow\pi}(0)$=0.26, $F_0^{B\rightarrow K}(0)$=0.29, $\gamma=110^0$
and $m_s(m_b)$=106 MeV.}
\vspace{1 cm}

\newpage
\centerline{ \DESepsf(dobphyspk0pi.epsf width 12 cm) }
\smallskip
\caption {Branching ratio for $B^{\pm} \rightarrow K^0 \pi^{\pm}$ as a function
of $\xi(\equiv {1\over N_c})$.
\newline The long dashed line :
$F_0^{B\rightarrow\pi}(0)$=0.26, $F_0^{B\rightarrow K}(0)$=0.29, $\gamma=110^0$
and $m_s(m_b)$=85 MeV.
\newline The dot-dash-dot line :
$F_0^{B\rightarrow\pi}(0)$=0.33, $F_0^{B\rightarrow K}(0)$=0.38, $\gamma=60^0$
and $m_s(m_b)$=85 MeV.
\newline The definitions for other lines are the same as those in Fig. 1.}
\vspace{1 cm}

\centerline{ \DESepsf(dobphypi1K1new1.epsf width 12 cm) }
\smallskip
\caption {Branching ratio for $B^0 \rightarrow K^{\pm} \pi^{\mp}$ as a function
of $\xi(\equiv {1\over N_c})$.
\newline The definitions for the lines are the same as those in Figs. 1 and 2.}
\vspace{1 cm}

\centerline{ \DESepsf(dobphypi0K1new1.epsf width 12 cm) }
\smallskip
\caption {Branching ratio for $B^{\pm} \rightarrow K^{\pm} \pi^0$ as a function
of $\xi(\equiv {1\over N_c})$.
\newline The definitions for the lines are the same as those in Figs. 1 and 2.}
\vspace{1 cm}

\centerline{ \DESepsf(dobphypi0K0new.epsf width 12 cm) }
\smallskip
\caption {Branching ratio for $B^0 \rightarrow K^0 \pi^0$ as a function of
$\xi(\equiv {1\over N_c})$.
\newline The definitions for the lines are the same as those in Figs. 1 and 2.}
\vspace{1 cm}

\centerline{ \DESepsf(dobphyspompi.epsf width 12 cm) }
\smallskip
\caption {Branching ratio for $B^{\pm} \rightarrow \omega \pi^{\pm}$ as a
function of $\xi(\equiv {1\over N_c})$.
\newline The solid line :
$F_0^{B\rightarrow\pi}(0)$=0.33, $F_0^{B\rightarrow K}(0)$=0.38, $\gamma=60^0$,
$A_0(0)=0.4$ and $m_s(m_b)$=106 MeV.
\newline The short dashed line :
$F_0^{B\rightarrow\pi}(0)$=0.33, $F_0^{B\rightarrow K}(0)$=0.38, $\gamma=110^0$,
$A_0(0)=0.4$ and $m_s(m_b)$=106 MeV.
\newline The dotted line :
$F_0^{B\rightarrow\pi}(0)$=0.33, $F_0^{B\rightarrow K}(0)$=0.38, $\gamma=60^0$,
$A_0(0)=0.36$ and $m_s(m_b)$=106 MeV.
\newline The dot-dashed line :
$F_0^{B\rightarrow\pi}(0)$=0.26, $F_0^{B\rightarrow K}(0)$=0.29, $\gamma=110^0$,
$A_0(0)=0.4$ and $m_s(m_b)$=106 MeV.}
\vspace{1 cm}

\centerline{ \DESepsf(dobphyspomk.epsf width 12 cm) }
\smallskip
\caption {Branching ratio for $B^{\pm} \rightarrow \omega K^{\pm}$ as a function
of $\xi(\equiv {1\over N_c})$.
\newline The long dashed line :
$F_0^{B\rightarrow\pi}(0)$=0.26, $F_0^{B\rightarrow K}(0)$=0.29, $\gamma=110^0$,
$A_0(0)=0.4$ and $m_s(m_b)$=85 MeV.
\newline The dot-dashed-dot line :
$F_0^{B\rightarrow\pi}(0)$=0.33, $F_0^{B\rightarrow K}(0)$=0.38, $\gamma=60^0$,
$A_0(0)=0.4$ and $m_s(m_b)$=85 MeV.
\newline The definitions for other lines are the same as those in Fig. 6.}
\vspace{1 cm}

\centerline{ \DESepsf(dobphyspomh1.epsf width 12 cm) }
\smallskip
\caption {Branching ratio for $B^{\pm} \rightarrow \omega h^{\pm}$ as a function
of $\xi(\equiv {1\over N_c})$.
\newline The definitions for the lines are the same as those in Figs. 6 and 7.}
\vspace{1 cm}

\centerline{ \DESepsf(dobphysphiknew.epsf width 12 cm) }
\smallskip
\caption {Branching ratio for $B^{\pm} \rightarrow \phi K^{\pm}$ as a function
of $\xi(\equiv {1\over N_c})$.
\newline The definitions for the lines are the same as those in Figs. 6 and 7.}
\vspace{1 cm}

\centerline{ \DESepsf(dobphysrhopinew.epsf width 12 cm) }
\smallskip
\caption {Branching ratios for $B^0 \rightarrow \rho^{\pm} \pi^{\mp}$ and
$B^{\pm} \rightarrow \rho^0 \pi^{\pm}$ as a function of $\xi(\equiv {1\over
N_c})$.
\newline The definitions for the lines are the same as those in Figs. 6 and 7.}
\vspace{1 cm}

\centerline{ \DESepsf(dobphyspkstrpi.epsf width 12 cm) }
\smallskip
\caption {Branching ratio for $B^0 \rightarrow K^{*\pm} \pi^{\mp}$ as a function
of $\xi(\equiv {1\over N_c})$.
\newline The definitions for the lines are the same as those in Figs. 6 and 7.}
\end{figure}

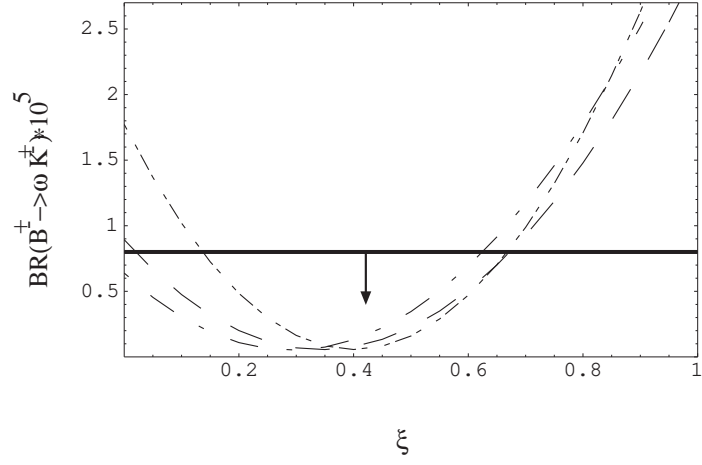
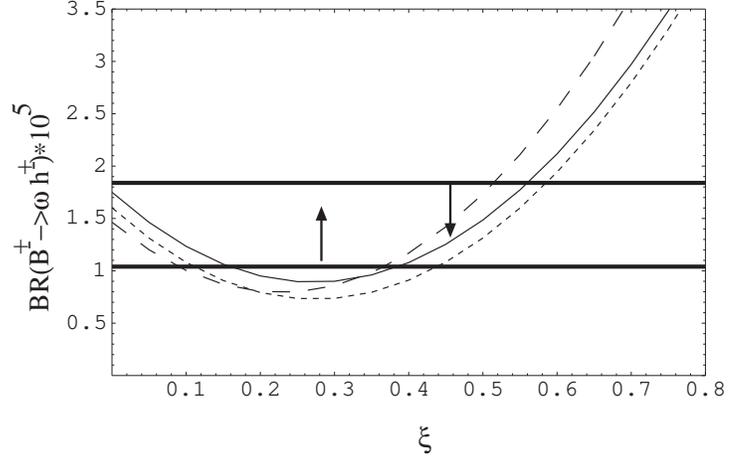
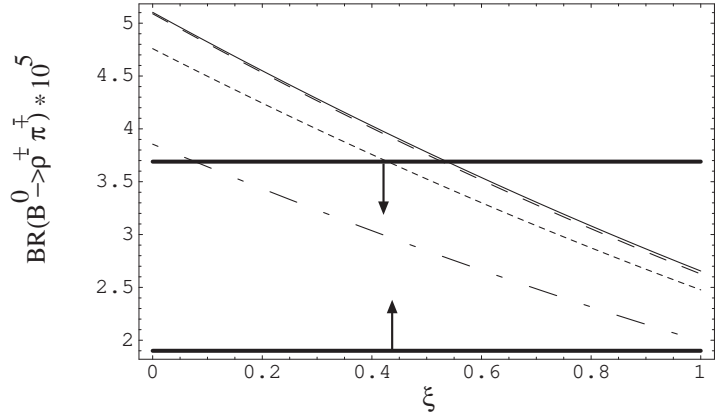
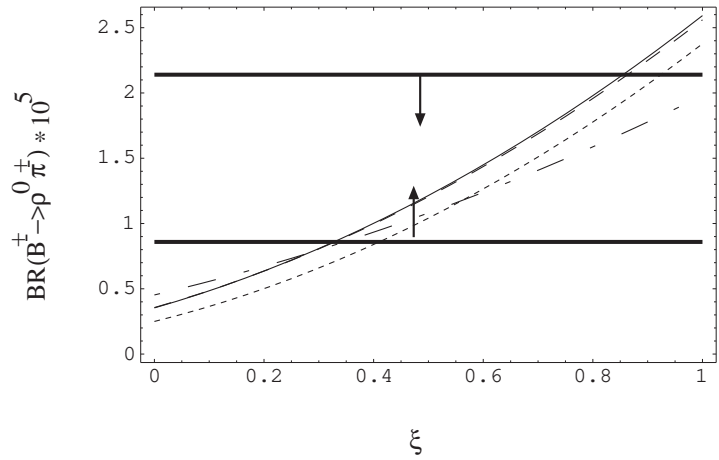
\end{document}